\documentclass[11pt]{article}

\usepackage[active]{srcltx}
\usepackage{hyperref}
\usepackage{verbatim}
\usepackage{amsmath,amssymb}
\usepackage[all]{xy}
\usepackage{graphicx}
\usepackage{color}
\usepackage{tikz-cd}
\usepackage{wasysym}
\usepackage{float}
 1 

\def\qed{\ifvmode\Realemovelastskip\fi
{\unskip\nobreak\hfil\penalty50\hbox{}\nobreak\hfil \hbox{\vrule
height1.2ex width1.2ex}\parfillskip=0pt \finalhyphendemerits=0
\par\smallskip}}
\def\qedr{\ifvmode\Realemovelastskip\fi
{\unskip\nobreak\hfil\penalty50\hbox{}\nobreak\hfil \hbox{
$\diamond$}\parfillskip=0pt \finalhyphendemerits=0
\par\smallskip}}

\newenvironment{proof}{\noindent{\sl Proof:~~~}}{\quad \qed}

\def\beq{\begin{equation}}
\def\eeq{\end{equation}}
\def\bea{\begin{eqnarray}}
\def\eea{\end{eqnarray}}
\def\beann{\begin{eqnarray*}}
\def\eeann{\end{eqnarray*}}
\def\beasn{\begin{sneqnarray}}
\def\eeasn{\end{sneqnarray}}
\def\ben{\begin{enumerate}}
\def\een{\end{enumerate}}
\def\bit{\begin{itemize}}
\def\eit{\end{itemize}}

\def\qed{\ifvmode\removelastskip\fi
{\unskip\nobreak\hfil\penalty50\hbox{}\nobreak\hfil \hbox{\vrule
height1.2ex width1.2ex}\parfillskip=0pt \finalhyphendemerits=0
\par\smallskip}}

\begin{document}

\centerline{\bf Reciprocal transformations and their role in the integrability and classification of PDEs}

\vskip 0.2cm
\vskip 0.5cm
\vskip 0.5cm

\centerline{P. Albares~$^a$, P. G. Est\'evez~$^a$ and C. Sard\'on~$^b$}
\vskip 0.5cm

\centerline{$^a$ Department of Theoretical Physics, University of Salamanca,   SPAIN.\\[10pt]}
\centerline{
$^b$  Instituto de Ciencias Matem\'aticas, Consejo Superior de Investigaciones Cient\'ificas.  Madrid. SPAIN\\
}
\vskip 0.5cm

\begin{abstract}
  \noindent
  Reciprocal transformations mix the role of the dependent and independent variables of  (nonlinear partial) differential equations to achieve simpler versions or even linearized versions of them. These transformations help in the identification of a plethora of partial differential equations that are spread out in the physics and mathematics literature.

Two different initial equations, although seemingly unrelated at first, could be the same equation after a reciprocal transformation. In this way, the big number of integrable equations that are spread out in the literature could be greatly diminished by establishing a method to discern which equations are disguised versions of a same, common underlying equation. Then, a question arises: Is there a way to identify different differential equations that are two different versions of a same equation in disguise?

\textbf{Keywords}: Reciprocal transformations, integrability, differential equations, water wave, Painlev\'e method, Painlev\' integrability
\end{abstract}

\section{Introduction}
\label{sec:The state of art}

On a first approximation, hodograph transformations are transformations involving
the interchange of dependent and independent variables \cite{clarkson,estevez05-1}. When the
variables are switched, the space of independent variables is called the reciprocal
space. In particular case of two variables, we refer to it as the reciprocal plane. As
a physical interpretation, whereas the independent variables play the role of positions
in the reciprocal space, this number is increased by turning certain fields or
dependent variables into independent variables and vice versa \cite{conte}. For example, in
the case of evolution equations in fluid dynamics, usually fields that represent the
height of the wave or its velocity, are turned into a new set of independent variables.
Reciprocal transformations share this definition with hodograph transformations, but these impose further requirements. Reciprocal transformations require the employment of
conservative forms together with the fulfillment their properties, as we shall see in forthcoming paragraphs
\cite{Est1,Est2,estevez05-1,{rogersshadwick}}. For example, some properties and requirements for reciprocal transformations that are not necessary for hodograph transformations are: the existence of conserved quantities for their construction \cite{Est1,Est2,EstSar2,EstSar1,rogers4,rogers5,rogerscarrillo}, that the invariance of certain integrable hierarchies under reciprocal
transformations induces auto-B\"acklund transformations \cite{EstSar2,EstSar1,oevelrogers, rogerscarrillo, rogersnucci}, and
these transformations map conservation laws to conservation laws and diagonalizable systems to diagonalizable systems, but act nontrivially on metrics and on Hamiltonian
structures.

But finding a proper reciprocal transformation is usually a very complicated
task. Notwithstanding, in fluid mechanics, a change of this type is usually
reliable, specifically for systems of hydrodynamic type. Indeed, reciprocal transformations have a long story alongside with the inverse scattering transform (IST) \cite{AbloClark,AbloSegur}, the two procedures gave rise to the discovery of other integrable nonlinear evolution equations similar to the KdV equations.  For example, Zakharov and Shabat \cite{ZS} presented the now famous nonlinear Sch\"odinger (NLS) equation, which presents an infinite number of integrals of motion and possesses $n$--soliton solutions with
purely elastic interaction. In 1928, the invariance of nonlinear gas dynamics, magnetogas dynamics and general hydrodynamic systems under reciprocal transformations
was extensively studied \cite{Ferapontov1,RogersKingstonShadwick}. Stationary and moving boundary problems in soil mechanics and nonlinear heat conduction
have likewise been subjects of much research \cite{FerapontovRogersSchief,Rogers11}.

One of the biggest advantages of dealing with hodograph and reciprocal transformations
is that many of the equations reported integrable in the bibliography of
differential equations, as the mentioned hydrodynamical systems, which are considered seemingly different from one another,
happen to be related via reciprocal transformations. If this were the case, two apparently
unrelated equations, even two complete hierarchies of partial differential equations (PDEs) that are linked
via reciprocal transformation, are tantamount versions of an unique problem. In this
way, the first advantage of hodograph and reciprocal transformations is that they give rise to a procedure of relating
allegedly new equations to the rest of their equivalent integrable sisters. The relation is achieved by finding simpler or linearized versions of a PDE so it becomes more tractable. For example, reciprocal transformations were proven to be a useful instrument to transform equations with peakon solutions into equations that are integrable in the Painlev\'e sense \cite{degasholm,h00}. Indeed, these transformations have also played an important role in  soliton theory and providing links between hierarchies of PDEs \cite{degasholm,h00}, as in relation to the aforementioned hydrodynamic-type systems. In this chapter we will depict straight forward reciprocal transformations that will help us identify different PDEs as different versions of a same problem,  as well as slight modifications of reciprocal transformations, as it can be compositions of several transformations of this type and others. For example, the composition of a {Miura transformation} \cite{AbloKruskalSegur,Sakovich} and a reciprocal transformation gives rise to the so called {Miura-reciprocal transformations} that helps us relate two different hierarchies of differential equations. A whole section of this chapter is devoted to illustrate Miura-reciprocal transformations.

A second significant advantage of reciprocal transformations is their
utility in the identification of integrable PDEs which a priori are not integrable
according to algebraic tests (for example, the Painlev\'e test is one of them)
\cite{Est2, estevez05-1} but they are proven indeed integrable according to Painlev\'e, after a reciprocal transformation. Our conjecture is that if an equation is integrable, there must be a
transformation that will let us turn the initial equation into a new one in
which the Painlev\'e test is successful. We will comment on this later in forthcoming paragraphs.

A third advantage for the use of reciprocal transformations is their role in the derivation of Lax pairs. Although it is not always possible to find a Lax pair for a given equation, a reciprocal transformation can turn it into a different one whose Lax pair is acknowledged. Therefore, by undoing the reciprocal transformation in the Lax pair of the transformed equation, we can achieve the Lax pair of the former.

These three main points describing the importance of reciprocal transformations imply the power of these transformations to classify differential equations and to sort out integrability.

\section{Fundamentals}

We will deal with some well-known differential equations in the literature of shallow water wave equations.
In particular, we will deal with generalizations of the Camassa--Holm equation and the Qiao equation \cite{CH,Est1,Est2,estevez05-1,qiao,qiao2007,QiaoLiu}. Such generalizations consist of a hierarchy, i.e., a set of differential equations that are related via a recursion operator. The recursive application of such operator gives members of different orders of the hierarchy, i.e., a set of different differential equations.
We will understand these differential equations as submanifolds of an appropiate higher-order tangent bundle.
Hence, let us introduce the necessary geometric tools for explaining PDEs as submanifolds of bundles.

\subsection{PDEs and jet bundles}
Let us consider a smooth $k$-dimensional manifold $N$ and the following projection $\pi:(x,u)\in \mathbb{R}^n\times N\equiv N_{\mathbb{R}^n}\mapsto x\in \mathbb{R}^n$ giving rise to a trivial bundle $(N_{\mathbb{R}^n}, \mathbb{R}^n,\pi)$. Here, we choose $\{x_1,\ldots,x_n\}$ as a global coordinate system on $\mathbb{R}^n$.

We say that two sections $\sigma_1,\sigma_2:\mathbb{R}^n\rightarrow N_{\mathbb{R}^n}$  are {$p$--equivalent at a point $x\in \mathbb{R}^n$} or they have a {contact of order $p$ at $x$}  if they have the same Taylor expansion of order $p$ at $x\in \mathbb{R}^n$. Equivalently,
\begin{equation}
\sigma_1(x)=\sigma_2(x),\qquad \frac{\partial^{|J|} (\sigma_1)_i}{\partial x_1^{j_1}\ldots\partial x_n^{j_n}}(x)=\frac{\partial^{|J|} (\sigma_2)_i}{\partial x_1^{j_1}\ldots\partial x_n^{j_n}}(x),
\end{equation}
for every multi-index $J=(j_1,\ldots,j_n)$ such that $0<|J|\equiv j_1+\ldots+j_n\leq p$ and $i=1,\dots,n$. Being $p$-equivalent induces an equivalence relation in the space $\Gamma(\pi)$ of sections of the bundle $(N_{\mathbb{R}^n},\mathbb{R}^n,\pi)$. Observe that if two sections have a contact of order $p$ at a point $x$, then they do have a contact at that point of the same type for any other coordinate systems on $\mathbb{R}^n$ and $N$, i.e., this equivalence relation is geometric.

We write $j_{x}^p\sigma$ for the equivalence class of sections that have a {contact of $p$-order} at $x\in \mathbb{R}^n$ with a section $\sigma$. Every such an equivalence
class is called a {$p$--jet}. We write ${\rm J}^{p}_x\pi$ for the space of all jets of order $p$ of sections at $x$. We will denote by ${\rm J}^p\pi$ the space of all jets of order $p$. Alternatively, we will write ${\rm J}^p(\mathbb{R}^n,\mathbb{R}^k)$ for the jet bundle of sections of the bundle $\pi:(x,u)\in\mathbb{R}^n\times\mathbb{R}^k\mapsto  x\in\mathbb{R}^n$.

Given a section $\sigma:\mathbb{R}^n\rightarrow {\rm J}^p\pi$, we can define the functions
\begin{equation}
(u_j)_J(j^p_x\sigma)=\frac{\partial^{|J|} \sigma_j}{\partial x_1^{j_1}\ldots\partial x_n^{j_n}}(x),\quad \forall j, \quad |J|\leq p.\end{equation} 
For $|J|=0$, we define $u_J(x)\equiv u(x)$. Coordinate systems on $\mathbb{R}^n$ and $N$ along with the previous functions give rise to a local coordinate system on ${\rm J}^p\pi$. We will also hereafter denote
the $n$-tuple and $k$-tuple, respectively, by $x=(x_1,\ldots,x_n),\,\, u=(u_1,\ldots,u_k)$, then
\begin{equation}\label{nose1}
(u_j)_J=u_{x_{i_1}^{j_1}\dots x_{i_n}^{j_n}}=\frac{\partial^{|J|} u_j}{\partial x_{i_1}^{j_1}\ldots \partial x_{i_n}^{j_n}},\quad \forall j,\quad |J|\leq 0.
\end{equation}

All such local coordinate systems give rise to a manifold structure on ${\rm J}^p\pi$. In this way, every point of ${\rm J}^p\pi$ can be written as
\begin{equation}\label{loccord}
\left(x_i,u_j,(u_j)_{x_i},(u_j)_{x_{i_1}^{j_1}x_{i_2}^{2-j_1}},(u_j)_{x_{i_1}^{j_1}x_{i_2}^{j_2}x_{i_3}^{3-j_1-j_2}},\dots,(u_j)_{x_{i_1}^{j_1}x_{i_2}^{j_2}\dots x_{i_n}^{p-\sum_{i=1}^{n-1}j_i}}\right),
\end{equation} 
where the numb indices run $i_1,\ldots,i_p=1,\dots,n$, $j=1,\dots,k,$ $j_1+\dots+j_n\leq p$. 

For small values of $p$, jet bundles have simple descriptions: ${\rm J}^{0}\pi=N_{\mathbb{R}^n}$ and ${\rm J}^1\pi\simeq \mathbb{R}^n\times {\rm T}N$.

The projections $\pi_{p,l}:j^p_x\sigma\in {\rm J}^p\pi\mapsto j^l_x\sigma\in {\rm J}^l\pi$ with $l<p$ lead to define the smooth bundles $({\rm J}^p\pi,{\rm J}^l\pi,\pi_{p,l})$.  
Conversely, for each section $\sigma: \mathbb{R}^n\rightarrow N_{\mathbb{R}^n}$, we have a natural embedding $j^p\sigma:\mathbb{R}^n\ni x\mapsto j^{p}_x\sigma \in {\rm J}^p\pi$. 

The differential equations that will be appearing along the chapter will be differential equations in close connection with shallow water wave models. We will define these PDEs
on a submanifold $N_{\mathbb{R}^n}$ of a higher-order bundle $J^p(\mathbb{R}^{n+1}, \mathbb{R}^{2k})$. For the reciprocal transformation, we will have to make use of conservation laws. By {conservation law} we will understand an expression of the form
\begin{equation}
\frac{\partial \psi_1}{\partial x_{i_1}}+\frac{\partial \psi_2}{\partial x_{i_2}}=0, 
\end{equation}
for certain two values of the indices in between $1 \leq i_1,i_2\leq n$ and two scalar functions $\psi_1,\psi_2 \in C^{\infty}({\rm J}^pN_{\mathbb{R}^n})$. The scalar fields representing water wave models will generally be denoted by $U$ or $u$, which depend on the independent variables $x_i$, and the functions $\psi_1,\psi_2$ will be functions of higher-order derivatives of $U$ or $u$. So, let us introduce the pairs $(u_j,x_i)$ or $(U_j,X_i)$ as local coordinates on the product manifold $N_{\mathbb{R}^n}$ and for the further higher-order derivatives we consider the construction given in \eqref{loccord}. In cases of lower dimensionality, as the 2-dim. case, we shall use upper/lower case  $(X,T)/(x,t)$. In the 3-dim. case, the independent variables will be denoted by upper/lower case $(X,T,Y)/(x,t,y)$.

\subsection{The Camassa--Holm hierarchy}

Let us consider the well-known Camassa--Holm equation (CH equation) in $1+1$ dimensions
 as a submanifold of $J^3(\mathbb{R}^2,\mathbb{R})$ with local coordinates for $\mathbb{R}^2\times \mathbb{R}$ the triple $(X,T,U)$. It reads:
\begin{equation}\label{cheq}
U_{T}+2\kappa U_{X}-U_{XXT}+3UU_{X}=2U_{X}U_{XX}+UU_{XXX}.
\end{equation}

We can interpret $U$ as the fluid velocity and $(X,T)$ as the spatial and temporal coordinates, respectively. Nonetheless, the equation \eqref{cheq} in its present form is not integrable in the strict defined Painlev\'e sense, but there exists a change of variables (action-angle
variables) such that the evolution equation in the new variables is equivalent to a
linear flow at constant speed. This change of variables is achieved by studying its
associated spectral problem and it is reminiscent of the fact that integrable classical
hamiltonian systems are equivalent to linear flows on tori, \cite{const2, const3, const1}. Indeed, \eqref{cheq} is a
bi-Hamiltonian model for shallow water waves propagation introduced by Roberto Camassa and Darryl Holm \cite{CH}. For $\kappa$ positive, the solutions are smooth solitons and for $\kappa=0$, it has peakon (solitons with a sharp peak, so with a discontinuity at the peak in the wave slope) solutions. A peaked solution is of the form:
\begin{equation}
U=ce^{-|X-cT|}+O(\kappa\log{\kappa}).
\end{equation}
In the following, we will consider the limiting case corresponding to $\kappa =0$. 

We can show the bi--Hamiltonian character of the equation by introducing the momentum
$M=U-U_{XX}
$, to write the two compatible Hamiltonian descriptions of the CH equation:
\begin{equation}
 M_{T}=-{\mathcal {D}}_{1}{\frac {\delta {\mathcal {H}}_{1}}{\delta M}}=-{\mathcal {D}}_{2}{\frac {\delta {\mathcal {H}}_{2}}{\delta M}},
\end{equation}
where
\begin{align}& {\mathcal {D}}_{1}=M{\frac {\partial }{\partial X}}+{\frac {\partial }{\partial X}}M,&{\mathcal {H}}_{1}={\frac {1}{2}}\int U^{2}+\left(U_{X}\right)^{2}\;{\text{d}}X,\nonumber\\
& {\mathcal {D}}_{2}={\frac {\partial }{\partial X}}-{\frac {\partial ^{3}}{\partial X^{3}}}, & {\mathcal {H}}_{2}={\frac {1}{2}}\int U^{3}+U\left(U_{X}\right)^{2}
{\text{d}}X.
\end{align}
The CH equation \eqref{cheq} is the first member of the well-known negative Camassa–-Holm hierarchy for a field $U(X,T)$ \cite{HolmQiao}. From now on, we will refer to this hierarchy by CH(1+1).

The CH(1+1) can be written in a compact form in terms of a recursion operator $R$, defined as follows:
\begin{equation}
U_T=R^{-n}U_X, \quad\quad\quad R=KJ^{-1},
\end{equation}
where $K$ and $J$ are defined as \begin{eqnarray}
&K=\partial_{XXX}-\partial_{X},\quad\quad\quad J=-\frac{1}{2}(\partial_XU+U\partial_X),\quad \partial_X=\frac{\partial}{\partial X}.\end{eqnarray} The factor $-\frac{1}{2}$ has
been conveniently added for future calculations.
We can include auxiliary fields $\Omega^{(i)}$ with $i=1,\dots,n$ when
the inverse of an operator appears. These auxiliary fields are defined as follows
\begin{align}
U_T&=J\Omega^{(1)},\nonumber\\
K\Omega^{(i)}&=J\Omega^{(i+1)},\quad i=1,\dots,n-1,\label{chh}\\
U_X&=K\Omega^{(n)}.\nonumber
\end{align}

It is also useful to introduce the change $U=P^2$, such that the final equations read:
\begin{align}
P_T&=-\frac{1}{2}\left(P\Omega^{(1)}\right)_X, \label{chcf}\\
\Omega^{(i)}_{XXX}-\Omega^{(i)}_{X}&=-P\left(P\Omega^{(i+1)}\right)_X,\quad i=1,\dots,n-1\label{cht1}\\
P^2&=\Omega^{(n)}_{XX}-\Omega^{(n)}.\label{cht2}
\end{align}

As we shall see in section 3, the conservative form of equation (\ref{chcf}) is the key for the study of reciprocal transformations.

\subsection{The Qiao hierarchy}

Qiao and Liu \cite{QiaoLiu} proposed an integrable equation defined as a submanifold of the bundle $J^3(\mathbb{R}^2,\mathbb{R})$. Notice that here the dependent variable is denoted by lower case $u$. In the future, we shall use lower cases for the dependent and independent variables related to Qiao hierarchy. The capital cases shall be used for Camassa--Holm. 
\begin{equation}\label{qiaoeq}
u_t=\left(\frac{1}{2u^2}\right)_{xxx}-\left(\frac{1}{2u^2}\right)_{x},
\end{equation}
\noindent
which also possesses peaked solutions as the CH equation, and a bi--Hamiltonian structure given by the relation

\begin{equation}
u_t=j\frac{\delta h_1}{\delta u}=k\frac{\delta h_2}{\delta u},
\end{equation}
where the operators $j$ an $k$ are
\begin{equation}
j=-\partial_xu\left(\partial_x\right)^{-1}u\partial_x,\qquad k=\partial_{xxx}-\partial_x,\quad \partial_x=\frac{\partial}{\partial x},
\end{equation}
and the Hamiltonian functions $h_1$ and $h_2$ correspond with
\begin{equation}
h_1=-\frac{1}{2}\int{\left[\frac{1}{4u^3}+\left(\frac{4}{5\,u^5}+\frac{4}{7\,u^7}\right)\,u_x^2\right]\,dx},\qquad h_2=-\int {\frac{1}{2u}\,dx}.
\end{equation}

We can define a recursion operator
as
\begin{equation}
r=kj^{-1}. 
\end{equation}
This recursion operator was used by Qiao in \cite{qiao2007} to construct a
$1+1$ integrable hierarchy, henceforth denoted as Qiao(1+1). This hierarchy reads

\begin{equation}
u_t=r^{-n}u_x.
\end{equation}
Equation \eqref{qiaoeq} is the second positive member of the Qiao hierarchy. The second negative member of the hierarchy was investigated by the same author in \cite{qiao}.
 If we introduce $n$ additional fields $v^{(i)}$ when
we encounter the inverse of an operator, the expanded equations read:
\begin{align}
u_t&=jv^{(1)},\nonumber\\
kv^{(i)}&=jv^{(i+1)},\quad i=1,\dots,n-1,\label{qh}\\
u_x&=kv^{(n)}.\nonumber
\end{align}

If we now introduce the definition of the operators $k$ and $j$, we obtain the following equations:
\begin{align}
u_t&=-\left(u\omega^{(1)}\right)_x,\label{qcf}\\
v^{(i)}_{xxx}-v^{(i)}_x&=-\left(u\omega^{(i+1)}\right)_x,\quad i=1,\dots,n-1,\label{qt1}\\
u&=v^{(n)}_{xx}-v^{(n)},\label{qt2}
\end{align}
in which $n$ auxiliary fields $\omega^{(i)}$ have necessarily been included to operate with the inverse term present in $j$. These fields have been defined as: \begin{equation}
\omega^{(i)}_x=uv^{(i)}_x, \quad i=1,\dots,n.    
\end{equation}
The conservative form of (\ref{qcf}) will allows us to define the reciprocal transformation.

\section{Reciprocal transformations as a way to identify and classify PDEs}
The CH(1+1) presented in the previous section is here explicitly shown to be equivalent
to $n$ copies of the Calogero-Bogoyavlenski-Schiff (CBS) equation \cite{bog,cal,pick}. This CBS equation
possesses the Painlev\'e property and the singular manifold method can be applied to obtain its Lax pair and
other relevant properties \cite{estevez05-1}.
Alongside, in the previous section we have also presented another example, the Qiao(1+1) hierarchy,  for which the Painlev\'e test
is neither applicable nor constructive. Nonetheless, here we will prove that there exists a reciprocal transformation which allows us to transform this hierarchy into $n$ copies of the modified Calogero-Bogoyavlenskii-Schiff
(mCBS), which is known to have the Painlev\'e property \cite{EstSar1}.
We shall denote the Qiao(1+1) likewise as mCH(1+1) because it can be considered as a modified version of the CH(1+1) hierarchy introduced in \cite{estevez05-1}.
Then, this subsection shows how different pairs of hierarchies and equations: CH(1+1) and CBS equation and the Qiao(1+1) or mCH(1+1) and mCBS equation are different versions of a same problem when a reciprocal transformation is performed upon them. Let us illustrate this in detail. 
\subsection{Hierarchies in $1+1$ dimensions}

\subsubsection*{Reciprocal transformations for CH(1+1)}

Given the conservative form of equation \eqref{chcf}, the following transformation arises naturally:
\begin{equation}\label{rtch}
dz_0=PdX-\frac{1}{2}P\Omega^{(1)}dT, \quad dz_1=dT.
\end{equation}
We shall now propose a reciprocal transformation \cite{EstSar1} by considering the former independent variable $X$ as a dependent field
of the new pair of independent variables $X=X(z_0,z_1),$ and therefore, $dX=X_0\,dz_0+X_1\,dz_1$ where the subscripts
zero and one refer to partial derivative of the field $X$ with respect to $z_0$ and $z_1$, correspondingly.
The inverse transformation takes the form:
\begin{equation}\label{irtch}
dX=\frac{dz_0}{P}+\frac{1}{2}\Omega^{(1)}dz_1,\quad dT=dz_1,
\end{equation}
which, by direct comparison with the total derivative of the field $X$, we obtain:
\begin{equation}
\partial_0X=\frac{\partial X}{\partial z_0}=\frac{1}{P},\quad \quad \partial_1X=\frac{\partial X}{\partial z_1}=\frac{\Omega^{(1)}}{2}.
\end{equation}
The important point  \cite{EstSar2,EstSar1} is that, we can now extend the transformation (\ref{rtch}) by introducing $n-1$ additional  independent variables $z_2,\dots,z_n$ which account for the transformation
of the auxiliary fields $\Omega^{(i)}$ in such a way that \begin{equation}  \quad\quad\quad \partial_iX=\frac{\partial X}{\partial z_i}=\frac{\Omega^{(i)}}{2},\quad \quad\quad i=2,\dots,n.\end{equation} Then, $X$ is a function
 $X=X(z_0,z_1,z_2,\dots,z_n)$ of $n+1$ variables.
It requires some computation to transform the hierarchy (\ref{chcf})-(\ref{cht2}) into the equations that  $X=X(z_0,z_1,z_2,\dots,z_n)$ should obey. For this matter, we use the symbolic calculus package Maple.
Equation \eqref{chcf} is identically satisfied by the transformation, and \eqref{cht1}, \eqref{cht2} lead to the following
set of PDEs:
\begin{equation}\label{bcbs}
\partial_0\left[-\frac{\partial_{i+1}X}{\partial_0X}\right]=\partial_i\left[\partial_0\left(\frac{\partial_{00}X}{\partial_0X}+\partial_0X\right)-\frac{1}{2}\left(\frac{\partial_{00}X}{\partial_0X}+\partial_0X\right)^2\right],\quad i=1,\dots,n-1,
\end{equation}
which constitutes $n-1$ copies of the same system, each of which is written in three variables $z_0,z_i,z_{i+1}$.
Considering the conservative form of \eqref{bcbs}, we shall introduce the change:
\begin{align}
\partial_iM&=\frac{1}{4}\left[-\frac{\partial_{i+1}X}{\partial_0X}\right],\\
\partial_0M&=\frac{1}{4}\left[\partial_0\left(\frac{\partial_{00}X}{\partial_0X}+\partial_0X\right)-\frac{1}{2}\left(\frac{\partial_{00}X}{\partial_0X}+\partial_0X\right)^2\right],\label{cbs2}
\end{align}
with $M=M(z_0,z_i,z_{i+1})$ and $i=1,\dots,n-1$.
The compatibility condition of $\partial_{000} X$ and $\partial_{i+1} X$ in this system gives rise to a set of  equations written entirely in terms of $M$:
\begin{equation}\label{cbs}
\partial_{0,i+1}M+\partial_{000i}M+4\partial_{i}M\partial_{00}M+8\partial_{0}M\partial_{0i}M=0,\quad i=1,\dots,n-1,
\end{equation}
which are  $n-1$ CBS equations \cite{bog, EstPrada, EstSar1}, each one in just three variables, for the field $M=M(z_0,..,z_i,z_{i+1},...z_n)$.

\subsubsection*{Reciprocal transformations for mCH(1+1)}
Given the conservative form of \eqref{qcf}, the following  reciprocal transformation \cite{EstSar1} naturally arises:
\begin{equation}\label{rtq}
dz_0=u\,dx-u\omega^{(1)}dt,\quad dz_1=dt.
\end{equation}
 We now propose a reciprocal transformation \cite{EstSar1} by considering the initial independent
variable $x$ as a dependent field of the new independent variables such that $x=x(z_0,z_1)$, and therefore, $dx=x_0\,dz_0+x_1\,dz_1$.
The inverse transformation adopts the form:
\begin{equation}\label{irtq}
dx=\frac{dz_0}{u}+\omega^{(1)}dz_1,\quad dt=dz_1.
\end{equation}
By direct comparison of the inverse transform with the total derivative of $x$, we obtain that:
\begin{equation}
\partial_0x=\frac{\partial x}{\partial z_0}=\frac{1}{u},\quad\quad \partial_1 x=\frac{\partial x}{\partial z_1}=\omega^{(1)}.
\end{equation}
We shall prolong this transformation in such a way that we introduce new variables $z_2,\dots,z_n$ such that $x=x(z_0, z_1,...,z_n)$ according to the following rule: \begin{equation}
\partial_i x=\frac{\partial x}{\partial z_i}=\omega^{(i)}, \quad \quad \quad i=2,\dots,n.\end{equation}
In this way, \eqref{qcf} is identically satisfied by the transformation, and \eqref{qt1}, \eqref{qt2} are transformed into $n-1$ copies
of the following equation, which is written in terms of just three variables $z_0,z_i,z_{i+1}$:
\begin{equation}\label{bmcbs}
\partial_0\left[\frac{\partial_{i+1}x}{\partial_0x}+\frac{\partial_{00i}x}{\partial_0x}\right]=\partial_i\left[\frac{(\partial_0x)^2}{2}\right],\quad i=1,\dots,n-1,
\end{equation}
The conservative form of these equations allows us to write them in the form of a system as:
\begin{align}
\partial_0\,m&=\frac{(\partial_0x)^2}{2},\label{mcbs1}\\
\partial_i\,m&=\frac{\partial_{i+1}x}{\partial_0x}+\frac{\partial_{00i}x}{\partial_0x},\quad i=1,\dots,n-1.\label{mcbs2}
\end{align}
which can be considered as  modified versions of the CBS equation with $m=m\left(z_0,..z_i,z_{i+1},...z_n\right)$. The modified CBS equation has been extensively studied from the point of view
of the Painlev\'e analysis in \cite{EstPrada}, its Lax pair was derived and hence, a version of a Lax pair for Qiao(1+1) is available in \cite{Est2,EstSar1}.

\subsection{Generalization to $2+1$ dimensions}
\subsubsection*{Reciprocal transformations for CH(2+1)}
From now on we will refer to the Camassa--Holm hierarchy in $2+1$ dimensions
as CH(2+1), and we will write it in a compact form as:
\begin{equation}
U_T=R^{-n}U_Y, \label{3.16}
\end{equation}
where $R$ is the recursion operator defined as:
\begin{equation}
R=JK^{-1},\quad K=\partial_{XXX}-\partial_X,\quad J=-\frac{1}{2}\left(\partial_XU+U\partial_X\right),\quad \partial_X=\frac{\partial}{\partial X}.
\label{3.17}
\end{equation}

This hierarchy was introduced in \cite{estevez05-1} as a generalization of the Camassa--Holm hierarchy. The recursion operator is the same as for CH(1+1). From this point of view, the spectral problem is the same \cite{calogero} and the $Y$-variable is just another ``time" variable  \cite{h00,ivanov}.

The $n$ component of this hierarchy can also be written as a set of PDEs by introducing $n$ dependent fields
$\Omega^{[i]}, (i=1\dots n)$ in the following way
\begin{eqnarray}
&&U_Y=J\Omega^{[1]}\nonumber\\&&J\Omega^{[i+1]}=K\Omega^{[i]},\quad i=1\dots n-1,\nonumber\\ \quad &&U_T=K\Omega^{[n]}, \label{3.18}
\end{eqnarray}
and by introducing two new fields, $P$ and $\Delta$, related to $U$ as:
\begin{equation}
U=P^2,\quad\quad P_T=\Delta_X, \label{3.19}
\end{equation}
we can write the hierarchy in the form of the following set of equations
\begin{eqnarray} &&P_Y=-\frac{1}{2}\left(P\Omega^{[1]}\right)_X,\nonumber\\
&&\Omega^{[i]}_{XXX}-\Omega^{[i]}_X=-P\left(P\Omega^{[i+1]}\right)_X,\quad i=1\dots n-1, \nonumber \\&& P_T=\frac{\Omega^{[n]}_{XXX}-\Omega^{[n]}_X}{2P}=\Delta_X.\label{3.20}
\end{eqnarray}

The conservative form of the first and third equation allows us to define the following exact derivative
\begin{equation}
dz_0= P\,dX-\frac{1}{2}P\Omega^{[1]}\,dY+\Delta\,dT. \label{3.21}
\end{equation}
A reciprocal transformation \cite{h00,rogers4,rogers5} can be introduced by considering the former independent variable $X$ as a field
depending on  $z_0$, $z_1=Y$ and $z_{n+1}=T$. From (\ref{3.21}) we have
\begin{eqnarray}
&&dX= \frac{1}{P}\,dz_0+\frac{\Omega^{[1]}}{2}\,dz_1-\frac{\Delta}{P}\,dz_{n+1},\nonumber\\&&Y=z_1,\quad\quad T=z_{n+1},\label{3.22}
\end{eqnarray}
and therefore
\begin{eqnarray}
&&\partial_0X=\frac{1}{P},\nonumber\\&&\partial_1X=\frac{\Omega^{[1]}}{2},\nonumber\label{8}\\&&\partial_{n+1}X=-\frac{\Delta}{P},\label{3.23}
\end{eqnarray}
where $\partial_i X=\frac{\partial X}{\partial z_i}$. We can now extend the transformation by introducing a new
independent variable $z_i$ for each field $\Omega^{[i]}$ by generalizing (\ref{3.23}) as
\begin{equation} \partial_i X=\frac{\Omega^{[i]}}{2},\quad i=2\dots n.\label{3.24}\end{equation}
Therefore, the new field $ X=X(z_0,z_1,\dots z_n,z_{n+1})$ depends on $n+2$ independent variables, where each  of the former dependent fields $\Omega_i,\,(i=1\dots n)$ allows us to define a new dependent variable $z_i$ through  definition (\ref{3.24}).
It requires some calculation (see \cite{estevez05-1} for  details) but it can be proved that the reciprocal transformation (\ref{3.22})-(\ref{3.24}) transforms (\ref{3.20}) to the following set of $n$ PDEs:
\begin{equation}\partial_0\left[-\frac{\partial_{i+1}X}{\partial_0X}\right]=\left[\partial_0\left(\frac{\partial_{00}X}{\partial_0X}+\partial_0X\right)-\frac{1}{2}\left(\frac{\partial_{00}X}{\partial_0X}+\partial_0X\right)^2\right]_i,\quad i=1\dots n. \label{3.25}\end{equation}
Note that each equation depends on only three  variables $z_0, z_i, z_{i+1}$. This result generalizes the one found in \cite{h00} for the first component of the hierarchy.
The conservative form of (\ref{3.25}) allows us to define a field $M(z_0,z_1,\dots z_{n+1})$ such that
\begin{eqnarray}\partial_iM&=&\frac{1}{4}\left[-\frac{\partial_{i+1}X}{\partial_0X}\right]=-\frac{P\Omega^{[i+1]}}{8},\quad\quad i=1\dots n-1,\nonumber\\
\partial_nM&=&\frac{1}{4}\left[-\frac{\partial_{n+1}X}{\partial_0X}\right]= \frac{\Delta}{4},\label{3.26}\\ \partial_0M&=&\frac{1}{4}\left[\partial_0\left(\frac{\partial_{00}X}{\partial_0X}+\partial_0X\right)-\frac{1}{2}\left(\frac{\partial_{00}X}{\partial_0X}+\partial_0X\right)^2\right]=\frac{1}{4P^2}\left(\frac{3P_X^2}{2P^2}-\frac{P_{XX}}{P}-\frac{1}{2}\right)\nonumber.\end{eqnarray}
It is easy to prove that each $M_i$ should satisfy the following CBS  equation  \cite{calogero} on $J^4(\mathbb{R}^{n+2},\mathbb{R})$,
\begin{equation}\partial_{0,i+1}M+\partial_{000i}M+4\partial_{i}M\partial_{00}M+8\partial_{0}M\partial_{0i}M=0,\quad i=1\dots n. \label{3.27}\end{equation}
Hence, the CH(2+1) is equivalent to $n$ copies of a CBS equation \cite{bog, cal, EstPrada} written in three different independent variables $z_0,z_i,z_{i+1}$.

\subsection*{Reciprocal transformation for mCH(2+1)}

Another example to illustrate the role of reciprocal transformations in the identification of partial differential equations was introduced by one of us in \cite{Est2}, were the following $2+1$ hierarchy Qiao(2+1) or mCH(2+1) appears as  follows.
\begin{equation}
u_t=r^{-n}u_y, \label{3.28}
\end{equation}
where $r$ is the recursion operator, defined as:
\begin{equation}
r=kj^{-1},\quad k=\partial_{xxx}-\partial_x,\quad j=-\partial_x\,u\,(\partial_x)^{-1}\,u\,\partial_x ,\quad \partial_x=\frac{\partial}{\partial x},
\label{3.29}
\end{equation}
where $\partial_x=\frac{\partial}{\partial x}$.
 This hierarchy generalizes the one introduced by Qiao in \cite{qiao2007}.
We shall  briefly summarize the results of \cite{Est2} when a  procedure similar to the one described above for CH(2+1) is applied to mCH(2+1).

If we introduce $2n$ auxiliary fields $v^{[i]}$, $\omega^{[i]}$ defined through
\begin{eqnarray}
&& u_y=jv^{[1]},\nonumber\\ &&jv^{[i+1]}=kv^{[i]},\quad \omega_x^{[i]}=uv_x^{[i]},\quad i=1\dots n-1,\nonumber\\ \quad && u_t=kv^{[n]}, \label{3.30}
\end{eqnarray}
the hierarchy can be expanded to $J^3(\mathbb{R}^3,\mathbb{R}^{2n+1})$
in the following form:
\begin{eqnarray}
 &&u_y=-\left(u\omega^{[1]}\right)_x,\nonumber\\&&
v^{[i]}_{xxx}-v^{[i]}_x=-\left(u\omega^{[i+1]}\right)_x,\quad i=1\dots n-1,\nonumber \label{18}\\&&u_t=\left(v^{[n]}_{xx}-v^{[n]}\right)_x,\label{3.31}
\end{eqnarray}
which allows us to define the exact derivative
\begin{equation}
dz_0= u\,dx-u\omega^{[1]}\,dy+\left(v^{[n]}_{xx}-v^{[n]}\right)\,dt \label{3.32}
\end{equation}
and $z_1=y, z_{n+1}=t$.
We can define a reciprocal transformation such that  the former independent variable $x$ is a new field $x=x(z_0,z_1,\dots \dots z_{n+1})$ depending on $n+2$ variables in the form
\begin{eqnarray}
&&dx=\frac{1}{u}dz_0+\omega^{[1]}dz_1-\frac{\left(v^{[n]}_{xx}-v^{[n]}\right)}{u}dz_{n+1},\nonumber\\&&y=z_1,\quad\quad t=z_{n+1},\label{3.33}
\end{eqnarray}
which implies
\begin{eqnarray}
&&\partial_0x=\frac{\partial x}{\partial z_0}=\frac{1}{u},\nonumber\\&&\partial_ix=\frac{\partial x}{\partial z_i}=\omega^{[i]},\quad i=1...n\label{20}\nonumber\\&&\partial_{n+1}x=\frac{\partial x}{\partial z_{n+1}}=-\frac{\left(v^{[n]}_{xx}-v^{[n]}\right)}{u}.\label{3.34}
\end{eqnarray}
The transformation of the equations (\ref{3.31}) yields the system of equations
\begin{equation}\partial_0\left[\frac{\partial_{i+1}x}{\partial_0x}+\frac{\partial_{00i}x}{\partial_0x}\right]=\partial_i\left[\frac{x_0^2}{2}\right],\quad i=1\dots n. \label{3.35}\end{equation}
Note that each equation depends on only three variables: $z_0, z_i, z_{i+1}$.

The conservative form of (\ref{3.35}) allows us to define a field $m=m(z_0,z_1,\dots z_{n+1})$ such that
\begin{eqnarray} &&\partial _0 m=\frac{x_0^2}{2}=\frac{1}{2u^2},\nonumber\\&& \partial_i m=\frac{\partial_{i+1}x}{\partial_0x}+\frac{\partial_{00i}x}{\partial_0x}=v^{[i]},\quad i=1\dots n,\label{3.36}\end{eqnarray}
defined on $J^3(\mathbb{R}^{n+2},\mathbb{R}^2)$. 
Equation (\ref{3.35}) has been extensively studied from the point of view of  Painlev\'e analysis \cite{EstPrada} and it can be considered as the modified version of the CBS equation (\ref{3.27}). 

Hence, we have shown again that a reciprocal transformation has proven the equivalency between two hierarchies/equations (mCH(2+1)-mCBS) that although they are unrelated at first, they are merely two different description of a same common problem.

\subsection{Reciprocal transformation for a fourth-order nonlinear equation}

In \cite{EstGand,EstLeble}, we introduced a fourth-order equation in $2+1$-dimensions which has the form
\begin{equation}
\left(H_{x_1x_1x_2} + 3H_{x_2}H_{x_1}-\frac{k+1}{4}\frac {(H_{x_1x_2})^2}{H_{x_2}}\right)_{x_1}=H_{x_2x_3} \label{3.37}. 
\end{equation}
The two particular cases $k=-1$ \cite{EstLeble} and $k=2$ \cite{EstGand,EstPra2} are integrable and it was possible to derive their Lax pair using the
singular manifold method \cite{weiss}. Based on the results in \cite{EstGand,EstLeble}, we proposed a spectral problem of the
form:
\begin{align}
&\phi_{x_1x_1x_1}-\phi_{x_3} + 3H_{x_1}\phi_{x_1}-\frac{k-5}{2}\,H_{x_1x_1}\phi=0,\nonumber\\
&\phi_{x_1x_2}+H_{x_2}\phi +\frac{k-5}{6}\,\frac{H_{x_1x_2}}{H_{x_2}}\,\phi_{x_2}=0\label{3.38}.
\end{align}

We can rewrite \eqref{3.37} as the system:
\begin{align}
&H_{x_1x_1x_2}+3H_{x_2}H_{x_1}-\frac{k+1}{4}\frac{H^2_{x_1x_2}}{H_{x_2}}=\Omega, \nonumber \\
&\Omega_{x_1}=H_{x_2x_3}.\label{3.39}
\end{align}
\subsubsection{Reciprocal transformation I}
We can perform a reciprocal transformation of equations \eqref{3.39} by
proposing:
\begin{align}
&dx_1=\alpha(x,t,T)[dx-\beta(x,t,T)dt-\epsilon(x,t,T)dT],\nonumber\\
&x_2=t,\quad x_3=T.\label{3.40}
\end{align}
Under this reciprocal transformation the derivatives transform as
\begin{eqnarray}
&&\frac{\partial}{\partial x_1}=\frac{1}{\alpha}\frac{\partial}{\partial x},\nonumber\\
&&\frac{\partial}{\partial x_2}=\frac{\partial}{\partial t}+\beta\frac{\partial}{\partial x},\nonumber\\
&&\frac{\partial}{\partial x_3}=\frac{\partial}{\partial T}+\epsilon\frac{\partial}{\partial x}.\label{3.41}
\end{eqnarray}

The cross derivatives of \eqref{3.40} give rise to the equations:
\begin{equation}\label{3.42}
\alpha_t + (\alpha\beta)_x=0,\quad \alpha_T+ (\alpha\epsilon)_x=0,\quad \beta_T-\epsilon_t+ \epsilon \beta_x-\beta\epsilon_x=0.
\end{equation} 
If we select a transformation in the form for $\alpha$ such that
\begin{equation}\label{3.43}
H_{x_2}=\alpha(x,t,T)^k,
\end{equation}
this reciprocal transformation, when applied to the system (\ref{3.39}), yields 

\begin{eqnarray}\label{3.44}
&&H_{x_1}= \frac{1}{3}\left(\frac{\Omega}{\alpha^k}-k\frac{\alpha_{xx}}{\alpha^3}+(2k-1)\left(\frac{\alpha_x}{\alpha^2}\right)^2\right),\\ &&
\label{3.45}
\Omega_x=-k\alpha^{(k+1)}\epsilon_x. 
\end{eqnarray}

Furthermore, the compatibility condition $H_{x_2x_1}=H_{x_1x_2}$ between \eqref{3.43} and \eqref{3.44} yields
\begin{equation}\label{3.46}
\Omega_t=-\beta\,\Omega_x-k\,\Omega\beta_x + \alpha^{k-2}\left[-k\beta_{xxx}+(k-2)\beta_{xx}\,\frac{\alpha_x}{\alpha}+3k\alpha^k\alpha_x\right].
\end{equation}
Then, the equations \eqref{3.42}, \eqref{3.45} and \eqref{3.46} constitute the transformed equations for the original system \eqref{3.39}. 

Still, we can find a more suitable form for the transformed equations if we introduce the following definitions:
\begin{equation}\label{3.47}
A_1=\frac{k+1}{3},\quad A_2=\frac{2-k}{3},\quad M=\frac{1}{\alpha^3}. 
\end{equation}
In these parameters, the integrability condition $(k+1)(k-2)=0$ is translated into
\begin{equation}\label{3.48}
A_1\cdot A_2=0,\quad A_1+A_2=1.
\end{equation}
Using the definitions above, we can finally present the reciprocally transformed system as:
\begin{eqnarray}\label{3.49}
&&A_1\left[\Omega_t+\beta\,\Omega_x+2\Omega\beta_x+2\beta_{xxx}+2\frac{M_x}{M^2}
\right]+\nonumber\\&&\quad\quad\quad\quad + A_2\left[\Omega_t+\beta\Omega_x-\Omega\beta_x-M\beta_{xxx}-M_x\beta_{xx}-M_x\right]=0,\nonumber\\
&&A_1\left(\Omega_x+2\,\frac{\epsilon_x}{M}\right)+A_2(\Omega_x-\epsilon_x)=0,\nonumber\\
&&M_t=3M\beta_x-\beta M_x,\nonumber\\&& M_T=3M\epsilon_x-\epsilon M_x, \nonumber\\ && \beta_T-\epsilon_t+ \epsilon \beta_x-\beta \epsilon_x=0.
\end{eqnarray}
Furthermore, the reciprocal transformation can also be applied to the spectral problem \eqref{3.38}. After some direct calculations, we obtain
\begin{equation}
\begin{aligned}
&A_1\left[\psi_{xt}+\beta\psi_{xx}-\left(\beta_{xx}-\frac{1}{M}\right)\psi\right]\\
&\qquad\quad+A_2\left[\psi_{xt}+\beta\psi_{xx}+2\beta_x\psi_x+\left(\beta_{xx}+1\right)\psi\right]M^{\frac{2}{3}}=0,\\[.5cm]
&A_1\left[\psi_{T}-M\psi_{xxx}-\left(M\Omega-\epsilon\right)\psi_x\right]\\
&\qquad\quad+A_2\left[\psi_{T}-M\psi_{xxx}-2M_x\psi_{xx}-\left(M_{xx}+\Omega-\epsilon\right)\psi_x\right]M^{\frac{2}{3}}=0,\label{3.50}
\end{aligned}
\end{equation}

where we have set
\begin{equation}\phi(x_1,x_2,x_3)=M^{\frac{1-2k}{9}}\psi(x,t,T) \end{equation}
for convenience.

\subsubsection*{Reduction independent of $T$}

Let us show a reduction of the set \eqref{3.49}, by setting all the fields independent of $T$. This means that
$$\epsilon= 0,\quad \Omega_x=0\Rightarrow \Omega=V(t), $$
and the system \eqref{3.49} reduces to
\begin{align}
&A_1\left[V_t+2\left(V\beta+\beta_{xx}-\frac{1}{M}\right)_x
\right]+ A_2\left[V_t-\left(V\beta+M\beta_{xx}+M\right)_x\right]=0,\label{3.52}\\
&M_t=3M\beta_x-\beta M_x.\label{3.53}
\end{align}

\noindent
$\bullet$ {\bf Degasperis--Procesi equation}

\medskip

For the case $A_1=1$ and $A_2=0$, we can integrate \eqref{3.52}  as:
\begin{equation}
\beta_{xx}+V\beta+\frac{V_t}{2}x=\frac{1}{M}+q_0,\end{equation} which combined with \eqref{3.53} yields
\begin{equation}\left(\beta_{xx} + V\beta\right)_t+\beta \beta_{xxx}+3\beta_x\beta_{xx}+4V\beta \beta_x-3q_0\beta_x+\frac{1}{2}V_t(\beta_x+3\beta x)+\frac{x}{2}V_{tt}=0.
\end{equation}
For $q_0=0$ and $V=-1$, this system is the well-known Degasperis–-Procesi equation, \cite{degasholm}.

\medskip

\noindent
$\bullet$ {\bf Vakhnenko equation}

\medskip

For the case $A_1=0,A_2= 1$, we can integrate \eqref{3.52}  as:
\begin{align}
&V_tx-V\beta-M\beta_{xx}-M-q_0=0,
\end{align}
which combined with \eqref{3.53} provides, when $V=0$, the derivative of the  Vakhnenko equation, \cite{vakh},

\begin{equation}\left[\left(\beta_t+\beta\beta_x\right)_x+3\beta\right]_x=0.\end{equation}

\subsubsection{Reciprocal transformation II}

A different reciprocal transformation can be constructed using the changes

\begin{align}\label{3.58}
&dx_2=\eta(y,z,T)\left(dz-u(y,z,T)dy-\omega(y,z,T)dT \right),\nonumber\\
&x_1=y,\quad x_3=T.
\end{align}

The compatibility conditions for this transformation are
\begin{align}
&\eta_y+(u\eta)_z=0,\nonumber\\
&\eta_T+(\eta\omega)_z=0,\nonumber\\
&u_T-\omega_y-u\omega_z+\omega u_z=0.\label{3.59}
\end{align}

We select the transformation by setting the field $H$ as the new independent variable $z$:
\begin{equation}\label{3.60}
z=H(x_1,x_2,x_3)\rightarrow dz=H_{x_1}dx_1+H_{x_2}dx_2+H_{x_3}dx_3. 
\end{equation}
By direct comparison of \eqref{3.58} and \eqref{3.60}, we obtain
\begin{align}
&H_{x_2}(x_1,x_2,x_3)=\frac{1}{\eta(y=x_1,z=H,T=x_3)},\nonumber\\
&H_{x_1}(x_1,x_2,x_3)=u(y=x_1,z=H,T=x_3),\nonumber\\
&H_{x_3}(x_1,x_2,x_3)=\omega(y=x_1,z=H,T=x_3),
\end{align}
and the transformations of the derivatives are
\begin{eqnarray}
&& \frac{\partial}{\partial x_1}=\frac{\partial}{\partial y}+u\frac{\partial}{\partial z},\nonumber\\
&&
\frac{\partial}{\partial x_2}=\frac{1}{\eta }\frac{\partial}{\partial z},\nonumber\\
&&
\frac{\partial}{\partial x_3}=\frac{\partial}{\partial T}+\omega\frac{\partial}{\partial z}.\label{3.62}
\end{eqnarray}

With this definitions, we get the transformation of the system \eqref{3.39},  as:
\begin{align}\label{3.63}
&G=(u_y+uu_z)_z+3u-\frac{k+1}{4}u_z^2,\nonumber\\
&G_y=(\omega-u G)_z,
\end{align}
where $G(z,y,T)$ has been defined as $G=\eta\, \Omega$.

\subsubsection*{Reduction independent of $T$}

The reduction independent of $T$ can be obtained
by setting $\omega=0$. In this case, the system \eqref{3.63} contains the case $G=0$,
\begin{eqnarray}
(u_y+uu_z)_z+3u-\frac{k+1}{4}u_z^2=0.
\end{eqnarray}
When $k=-1$, it is the Vakhnenko equation. For the other integrable case, $k=2$, it yields a modified Vakhnenko equation if $A_2=0$.

\section{Reciprocal transformations to derive Lax pairs}

Reciprocal transformations have served us as a way to derive Lax pairs of differential equations and hierarchies of such differential equations. A differential equation in its initial form may not be Painlev\'e integrable as we mentioned before, but we are able to prove its integrability by transforming it into another differential equation via reciprocal transformation that makes it Painlev\'e integrable. In the same fashion, an initial differential equation may not have an associated Lax pair and the singular manifold method may not be applicable. Through a reciprocal transformation we can again transform such equation into another in which we can work the singular manifold method upon.
We are depicting examples in the following lines.
\subsection{Lax pair for the CH(2+1) hierarchy}
In section 2, we have proved that the reciprocal transformations can be used to establish the equivalence between the CH(2+1) hierarchy (\ref{3.16}) and $n+1$ copies of the CBS equation (\ref{3.27}). This CBS equation has the Painlev\'e property \cite {pick} and the singular manifold method can be successfully used to derive the following Lax pair \cite{EstPrada}, 

\begin{align}
&\partial_{00}\psi=\left(-2\partial_0 M-\frac{\lambda}{4}\right)\,\psi,\label{4.1}\\
&0=E_i=\partial_{i+1}\psi-\lambda\partial_i\psi+4\partial_iM\partial_0\psi-2\partial_{0i}M\,\psi\label{4.2}.\end{align}

Furthermore, the compatibility condition between these two equations implies that the spectral problem
is nonisospectral because $\lambda$ satisfies:
\begin{equation}
\partial_0\lambda=0, \quad\quad \partial_{i+1}\lambda-\lambda\partial_i\lambda=0. \label{4.3}
\end{equation}

Notice that the first equation in the Lax pair is independent of the index $i$. Nevertheless, the second equation can be considered as a recursion relation for the derivatives of $\psi$ with respect to each $z_i$.

Now, to come back to the original fields $U$ and $\Omega^{[i]}$ as well as to the original variables $X,Y,T$, all we need is to perform the change
\begin{equation}
\psi(z_0,z_1,\dots,z_n,z_{n+1})=\sqrt{P}\,\phi(X,Y,T)    \label{4.4}
\end{equation}
where $P$ is defined in (\ref{3.19}).
Considering the reciprocal transformation \eqref{3.22}, we have the following induced transformations
\begin{alignat}{3}
&\partial_0\psi&&=\sqrt{P}\left(\frac{\phi_X}{P}+\frac{P_X}{2P^2}\phi\right),\nonumber\\
&\partial_{00}\psi&&=\sqrt{P}\left(\frac{\phi_{XX}}{P^2}+\left[\frac{P_{XX}}{2P^3}-\frac{3}{4}\frac{P_X^2}{P^4}\right]\phi\right),\nonumber\\
&\partial_1\psi&&=\sqrt{P}\left(\phi_Y+
\frac{\Omega^{[1]}\phi_X}{2}+\left[\frac{P_Y}{2P}+\frac{P_X\Omega^{[1]}}{4P}\right]\phi\right)\nonumber\\
&\quad\quad&&=\sqrt{P}\left(\phi_Y+
\frac{\Omega^{[1]}\phi_X}{2}-\frac{\Omega^{[1]}_X\phi}{4}\right),\nonumber\\
&\partial_{n+1}\psi&&=\sqrt{P}\left(\phi_T-\frac{\Delta\phi_X}{P}+\left[\frac{P_T}{2P}-\frac{P_X\Delta}{2P^2}\right]\phi\right)\nonumber\\
&\quad\quad&&=\sqrt{P}\left(\phi_T-\frac{\Delta\phi_X}{P}+\left[\frac{\Delta_X}{2P}-\frac{P_X\Delta}{2P^2}\right]\phi\right).
\label{4.5}
\end{alignat}
With these changes, (\ref{4.1}) becomes: 
\begin{equation}
\phi_{XX}+\left(\frac{\lambda P^2}{4}-\frac{1}{4}\right)\phi=0,\nonumber \end{equation}
where equation (\ref{3.26}) has been used. Finally, the combination with  (\ref {3.19}) yields
\begin{equation} \phi_{XX}=\frac{1}{4}\left(1-\lambda U\right)\phi,\label{4.6}\end{equation}
as the spatial part of the Lax pair for the  CH(2+1) hierarchy. The temporal part can be obtained from (\ref{4.2}) through the following combination: 
\begin{equation}
0=\sum_{i=1}^n\lambda^{n-i}E_i=\sum_{i=1}^n\lambda^{n-i}\left(\partial _{i+1}\psi-\lambda\partial_i\psi\right)+\sum_{i=1}^n\lambda^{n-i}\left(4\partial_iM\partial_0\psi-2\partial_{0i}M\,\psi\right).\label{4.7}\end{equation}
It is easy to prove that 
\begin{equation}\sum_{i=1}^n\lambda^{n-i}\left(\partial _{i+1}\psi-\lambda\partial_i\psi\right)=\partial_{n+1}\psi-\lambda^n\partial_1\psi.\label{4.8}\end{equation}
The reciprocal transformation (\ref{3.22}), when applied to (\ref{4.8}), and combined with (\ref{4.4}) and (\ref{4.5}) yields

\begin{eqnarray}
&&\sum_{i=1}^n\lambda^{n-i}\left(\partial _{i+1}\psi-\lambda\partial_i\psi\right)=\sqrt{ P}\left[\phi_T-\frac{\Delta\phi_X}{P}+\frac{\Delta_X\phi}{2P}-\frac{P_X\Delta}{2P^2}\,\phi\right]\nonumber\\&&\quad\quad\quad\quad\quad-\,\lambda^n\,\sqrt{ P}\left[\phi_Y+
\frac{\Omega^{[1]}\phi_X}{2}-\frac{\Omega^{[1]}_X\phi}{4}\right].\label{4.9}
\end{eqnarray}
For the last sum of (\ref{4.7}), we can use \eqref{3.26} and (\ref{4.5}). The result is 
\begin{eqnarray}
&&\sum_{i=1}^n\lambda^{n-i}\left(4\partial_iM\partial_0\psi-2\partial_{0i}M\,\psi\right)=\sqrt P\left[\frac{\Delta\phi_X}{P}+\frac{\Delta\,P_X}{2P^2}\phi-\frac{\Delta_X}{2P}\phi\right]\nonumber\\&&\quad\quad\quad\quad+\sqrt{P}\, \sum_{i=1}^{n-1}\lambda^{n-i}\left[\frac{\Omega^{[i+1]}_X}{4}\phi-\frac{\Omega^{[i+1]}}{2}\phi_X\right]=0.\label{4.10}
\end{eqnarray}
Substitution of (\ref{4.9}) and (\ref{4.10}) in (\ref{4.7}) yields
\begin{equation}\phi_T-\lambda^n\phi_Y+\lambda^n\left(\frac{\Omega^{[1]}_X}{4}\phi-\frac{\Omega^{[1]}}{2}\phi_X\right)+\sum_{i=1}^{n-1}\lambda^{n-i}\left[\frac{\Omega^{[i+1]}_X}{4}\phi-\frac{\Omega^{[i+1]}}{2}\phi_X\right]=0.\label{4.11}\end{equation}

The expression (\ref{4.11}) can be written in a more compact form as 
\begin{equation}
\phi_T-\lambda^n\phi_Y+\frac{A_X}{4}\phi-\frac{A}{2}\phi_X=0,\label{4.12}\end{equation}
where $A$ is defined as
\begin{equation}A=\sum_{j=1}^n\lambda^{n-j+1}\Omega^{[j]}, \quad\quad \text{with}\quad i=j-1.\end{equation}

The nonisospectral condition (\ref{4.3}) reads
\begin{equation}\lambda_X=0,\quad\quad 0=\sum_{i=1}^n \lambda^{n-i}\left(\partial_{i+1}-\lambda\partial_i\right)\lambda=\partial_{n-1}\lambda-\lambda^n\,\partial_1\lambda=\lambda_T-\lambda^n\lambda_Y=0.\end{equation}

In sum: the Lax pair for CH(2+1) can be written as
\begin{eqnarray} &&\phi_{XX}+\frac{1}{4}\left(\lambda\,U-1\right)\phi=0,\nonumber\\&& \phi_T-\lambda^{n}\phi_Y -\frac{A}{2}\phi_X+\frac{A_X}{4}\phi=0,\label{4.15}\end{eqnarray}
where
\begin{equation} A=\sum_{i=1}^{n}
\left[\lambda^{n-i+1}\,\Omega^{[i]}\right],\qquad \lambda_T-\lambda^{n}\lambda_Y=0.\label{4.16}
\end{equation}

\subsection{Lax pair for mCH(2+1)}

In \cite{EstPrada} it was proved that the CBS equation (\ref{3.27}) and the mCBS equation (\ref{3.35}) were linked through a Miura transformation. This is a transformation that relates the fields in the CBS and mCBS in the following form
\begin{eqnarray} \partial_0M&=&-\frac{\partial_0x^2}{8}+ \frac{\partial_{00}x}{4},\nonumber\end{eqnarray} 
which combined with (\ref{3.35}) can be integrated as
\begin{eqnarray}  4M= \partial_0x-m.\label{4.17}\end{eqnarray}

The two-component Lax pair for the mCBS equation (\ref{3.35}) was derived in \cite {EstPrada}. In our variables this spectral problem reads:
\begin{equation}\partial_0 \left(
\begin{array}{c} \psi \\ \hat \psi
\end{array}
\right) =\frac{1}{2}\left( \begin{array}{cc} -\, \partial_0 x& i\sqrt {
\lambda}\\ i\sqrt {
\lambda}
& \partial_0 x
\end{array}
\right) \left(
\begin{array}{c} \psi \\ \hat \psi
\end{array}
\right),\label{4.18}\end{equation}

\begin{eqnarray}&&0=F_i=\partial_{i+1} \left(
\begin{array}{c} \psi \\ \hat \psi
\end{array}
\right) -\lambda\,\partial_{i} \left(
\begin{array}{c} \psi \\ \hat \psi
\end{array}
\right)\nonumber\\ &&\quad\quad\quad-\frac{1}{2}\left( \begin{array}{cc} -\, \partial_{i+1} x &i\sqrt {
\lambda}\,\partial_i\left(m-\, \partial_0 x\right) \\ i\sqrt {
\lambda}\,\partial_i\left(m+ \partial_0 x\right) &
\partial_{i+1} x
\end{array}
\right) \left(
\begin{array}{c} \psi \\ \hat \psi
\end{array}
\right).\label{4.19}\end{eqnarray}
  It is easy to see that the compatibility condition of (\ref{4.18})-(\ref{4.19}) yields the equation
  (\ref{3.35}) as well as the following nonisospectral contition:
\begin{equation} \partial_0\lambda=0,\quad\quad
\partial_{i+1}\lambda=\lambda\,\partial_i\lambda.\label{4.20}\end{equation}

If, from the above Lax pair, we wish to obtain the spectral problem of the mCH(2+1), we need  to invert
the reciprocal transformation (\ref{3.33})-(\ref{3.34}), which means applying the following substitutions:
\begin{eqnarray}
&&\partial_0x=\frac{1}{u},\nonumber\\&&\partial_ix=\omega^{[i]}\quad\Rightarrow\quad \partial_{0i}x=\frac{\omega^{[i]}_x}{u}=v^{[j]}_x, \quad\quad i=1...n,\nonumber\\
&&\partial_{n+1}x=-\frac{v^{[n]}_{xx}-v^{[n]}}{u},\label{4.21}\\&&\partial_0m=\frac{1}{2u^2},\nonumber\\&&\partial_im=v^{[i]}\nonumber.
\end{eqnarray}
and the transformations of the derivatives are

\begin{eqnarray}&&\partial_0=\frac{1}{u}\,\partial_x\,,\nonumber\\ &&\partial_1=\partial_y+\omega^{[1]}\,\partial_x\,,
\label{4.22}\\
&&\partial_{n+1}=\partial_t-\frac{\left(v^{[n]}_{xx}-v^{[n]}\right)}{u}\,\partial_x.\nonumber\end{eqnarray}

We can now tackle the transformation of the Lax pair (\ref{4.18})-(\ref{4.19}). The spatial part (\ref{4.18}) transforms trivially 
to:
\begin{equation} \left(
\begin{array}{c} \psi \\ \hat \psi
\end{array}
\right)_x =\frac{1}{2}\left( \begin{array}{cc} -\,1& i\sqrt {
\lambda}u\\ i\sqrt {
\lambda}u
&1
\end{array}
\right) \left(
\begin{array}{c} \psi \\ \hat \psi
\end{array}
\right).\label{4.23}\end{equation}

The transformation of (\ref{4.19})
is  slightly  more complicated. Let us compute the
following sum:
\begin{eqnarray}
0=\sum_{i=1}^{n}\lambda^{n-i}F_i,\label{4.24}
\end{eqnarray} 
where $F_i$ is defined in (4.19). It is easy to see that 
\begin{equation}\sum_{i=1}^n\lambda^{n-i}(\partial_{i+1}-\lambda\partial_i) \left(
\begin{array}{c} \psi \\ \hat \psi
\end{array}
\right) =(\partial_{n+1}-\lambda^n\partial_1) \left(
\begin{array}{c} \psi \\ \hat \psi
\end{array}
\right),\label{4.25}\end{equation}
and then, the inverse reciprocal transformation (\ref{4.21})-(\ref{4.22}) can be applied to (\ref{4.24}) in order to obtain
\begin{equation} \left(
\begin{array}{c} \psi \\ \hat \psi
\end{array}
\right)_t-\lambda^n\left(
\begin{array}{c} \psi \\ \hat \psi
\end{array}
\right)_y =C\left(
\begin{array}{c} \psi \\ \hat \psi
\end{array}
\right)_x+\frac{i\sqrt {
\lambda}}{2}\left( \begin{array}{cc} 0&B_{xx}-B_x\\ B_{xx}+B_x
&0
\end{array}
\right) \left(
\begin{array}{c} \psi \\ \hat \psi
\end{array}
\right),\label{4.26}\end{equation}
where 
\begin{equation}C=\sum_{i=1}^{n}\lambda^{n-i+1}\omega^{[i]},\quad\quad B=\sum_{i=1}^{n}\lambda^{n-i}v^{[i]}.\quad \label{4.27}\end{equation}
The inverse reciprocal transformation, when applied to (\ref{4.20}) yields
\begin{equation}\lambda_x=0,\quad\quad 0=\sum_{i=1}^n \lambda^{n-i}\left(\partial_{i+1}-\lambda\partial_i\right)\lambda=\partial_{n-1}\lambda-\lambda^n\,\partial_1\lambda=\lambda_t-\lambda^n\lambda_y=0.\end{equation}

Hence, we have derived a Lax pair for mCH(2+1) using the existing Miura transformation between the CBS and mCBS and the Lax pair for mCBS. This is another example of how reciprocal transformations or compositions of transformations can provide us with Lax pairs, and the implication of integrability.
\section{A Miura-reciprocal transformation}

Recalling the previous sections, we can summarize by saying CH(2+1) and mCH(2+1) are related to the CBS and mCBS by reciprocal transformations, correspondingly.
Aside from this property, in this section we would like to show that there exists a Miura transformation \cite{EstPrada} relating the CBS and the mCBS equations. Hence, one wonders if mCH(2+1) is related
to CH(2+1) in any way. It seems clear that the relationship between mCH(2+1) and CH(2+1) necessarily includes a composition of a Miura and a reciprocal transformation. 

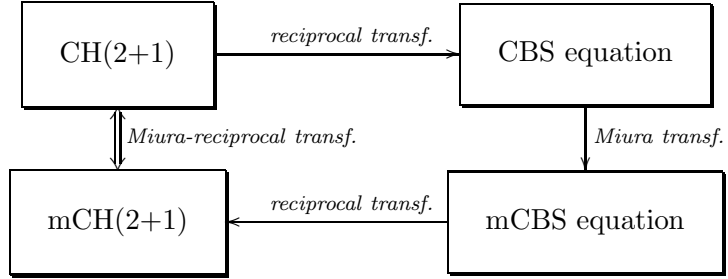
\begin{figure}[H]
\begin{center}
$
\xymatrix{*+<1cm>[F-,]{\text{CH(2+1)}} \ar[rrr]^{\textit{reciprocal transf.}} \ar@2{<->}[d]^{\textit{Miura-reciprocal transf.}} &  &  &*+<1cm>[F-,]{\text{CBS equation}}\ar[d]^{\textit{Miura transf.}}\\  *+<1cm>[F-,]{\text{mCH(2+1)}} &  &  &*+<1cm>[F-,]{\text{mCBS equation}}\ar[lll]_(0.5){\textit{reciprocal transf.}}}
$
\end{center}
\caption{Miura-reciprocal transformation.}
\label{Fig2}
\end{figure}

Evidently, the relationship between both hierarchies cannot be a simple Miura transformation because they are written in different variables $(X,Y,T)$ and $(x,y,t)$. The answer is provided by the relationship of both sets of variables with the same set $(z_0,z_1,z_{n+1})$. 
By combining (\ref{3.21}) and (\ref{3.32}), we have
\begin{equation}
\begin{aligned}
 &P\,dX-\frac{1}{2}P\Omega^{[1]}\,dY+\Delta\,dT=u\,dx-u\omega^{[1]}\,dy+\left(v^{[n]}_{xx}-v^{[n]}\right)\,dt,\\
&Y=y,\quad\quad\quad T=t, \label{5.1}
\end{aligned}\end{equation} 
which yields the required relation between the independent variables of CH(2+1) and those of mCH(2+1). 
The Miura transformation (\ref{4.17}), combined with (\ref{3.26}) and (\ref{3.36}) also provides the following results
\begin{eqnarray}
 &&4\partial_{0}M=\partial_{00}x-\partial_{0}m\Longrightarrow \frac{\partial_{00}X}{\partial_{0}X}+\partial_{0}X=\partial_{0}x, \label{5.2}\\ &&4\partial_{i}M=\partial_{0i}x-\partial_{i}m\Longrightarrow -\frac{\partial_{i+1}X}{\partial_{0}X}=\partial_{0i}x-\frac{\partial_{00i}x}{\partial_{0}x}-\frac{\partial_{i+1}x}{\partial_{0}x}, \label{5.3}\end{eqnarray} 
with $i=1,\dots, n$.
With the aid of (\ref{3.23}), (\ref{3.24}) and (\ref{3.34}), the following results 
 arise from (\ref{5.2})-(\ref{5.3}) 
\begin{eqnarray}
 &&\frac{1}{u}=\left(\frac{1}{P}\right)_X+\frac{1}{P},\nonumber\\&& P\Omega^{[i+1]}= 2\left(v^{[i]}-v^{[i]}_x\right)\quad\Longrightarrow\quad \omega^{[i+1]}=\frac{\Omega^{[i+1]}_X+\Omega^{[i+1]}}{2},\quad i=1\dots n-1,\nonumber\\&& \Delta =v^{[n]}_x-v^{[n]}.\label{5.4}
\end{eqnarray}

Furthermore, (\ref{5.1}) can be integrated as
\begin{equation}x=X-\ln P.\label{5.5}\end{equation}
By summarizing the above conclusions, we have proven that the mCH(2+1)  hierarchy
 \begin{equation}
u_t=r^{-n}u_y,\quad u=u(x,y,t), \end{equation}  can be considered as the modified version of CH(2+1)
  \begin{equation}
U_T=R^{-n}U_Y,\quad U=U(X,Y,T).  \end{equation} 
The transformation that connects the two hierarchies  involves the reciprocal transformation
\begin{equation}x=X-\frac{1}{2}\ln U,\label{5.8}\end{equation}
as well as the following  transformation between the fields
\begin{eqnarray}
&&\frac{1}{u}=\frac{1}{\sqrt U}\left(1-\frac{U_X}{2U}\right), \nonumber\\&& \omega^{[i]}=\frac{\Omega^{[i]}_X+\Omega^{[i]}}{2},\quad i=1\dots n,\nonumber\\&&
 \frac{\delta}{u}=\left(\frac{\Delta}{\sqrt U}\right)_X+\frac{\Delta}{\sqrt U}\label{5.9}.\end{eqnarray} 
\subsection{Particular case 1: The Qiao equation}
We are now restricted to the first component of the hierarchies $n=1$ in the case in which the field $u$ is independent of $y$ and $U$ is independent of $Y$.
\begin{itemize}
\item From (\ref{3.19}) and (\ref{3.20}), for the restriction of CH(2+1) we have
\begin{equation}
\begin{aligned}&U=P^2, \\&U_T=\Omega^{[1]}_{XXX}-\Omega^{[1]}_{X},\\& \left(P\,\Omega^{[1]}\right)_X=0,\label{5.10}\end{aligned}\end{equation} 
which can be summarized as
\begin{equation}
\begin{aligned}\Omega^{[1]}&=\frac{k_1}{P}=\frac{k_1}{\sqrt U},\\
U_T&=k_1\left[\left(\frac{1}{\sqrt U}\right)_{XXX}-\left(\frac{1}{\sqrt U}\right)_{X}\right],\label{5.11}\end{aligned}\end{equation} 
that is the {Dym equation} \cite{kruskal}.

\item The reduction of mCH(2+1)  can be achieved from \eqref{18} in the form
\begin{equation}
\begin{aligned}&\omega^{[1]}_x=uv^{[1]}_x,\\&u_t=v^{[1]}_{xxx}-v^{[1]}_{x},\\& \left(u\omega^{[1]}\right)_x=0,\end{aligned}\label{5.12}\end{equation} 
which can be written as
\begin{equation}\label{5.13}
\begin{aligned}&\omega^{[1]}=\frac{k_2}{u} \quad\Longrightarrow
\quad v^{[1]}=\frac{k_2}{2u^2},\\
&u_t=k_2\left[\left(\frac{1}{2u^2}\right)_{xx}-\left(\frac{1}{2u^2}\right)\right]_{x},\end{aligned}\end{equation} 
that is the Qiao equation.

\item From \eqref{5.8} and \eqref{5.9} it is easy to see that $k_1=2k_2$. By setting $k_2=1$, we can conclude that the Qiao equation
\begin{equation}
u_t=\left(\frac{1}{2u^2}\right)_{xxx}-\left(\frac{1}{2u^2}\right)_{x},\end{equation} 
is the modified version of the Dym equation
\begin{equation}
U_T=\left(\frac{2}{\sqrt U}\right)_{XXX}-\left(\frac{2}{\sqrt U}\right)_{X}.\end{equation} 
\item From \eqref{8} and \eqref{20}, it is easy to see that the independence from $y$ implies that $\partial_1X=\partial_0X$ and $\partial_1 x=\partial_0 x$, which means that  the CBS and mCBS \eqref{3.27} and \eqref{3.35} reduce to the following potential versions of the KdV and modified KdV equations
\begin{equation}
\begin{gathered} \partial_0\left(\partial_2 M+\partial_{000}M+6\partial_0M^2\right)=0,\\
\partial_2 x+\partial_{000}x-\frac{1}{2}\partial_0x^3=0.\end{gathered}\end{equation} 
\end{itemize}

\subsection{Particular case 2: The Camassa--Holm equation}
If we are restricted to the $n=1$ component when $T=X$ and $t=x$, the following results hold:
\begin{itemize}
\item From \eqref{3.19} and \eqref{3.20}, for the restriction of CH(2+1) we have
\begin{equation}
\begin{aligned} &\Delta=P=\sqrt U,\\& U=\Omega^{[1]}_{XX}-\Omega^{[1]},\\ &U_Y+U\Omega^{[1]}_X+\frac{1}{2}\Omega^{[1]} U_X=0,\end{aligned}\end{equation} 
which is the Camassa--Holm equation.

\item The reduction of mCH(2+1)  can be obtained from \eqref{18} in the form
\begin{equation}
\begin{aligned}&\delta=u=v^{[1]}_{xx}-v^{[1]},\\& u_y+\left(u\omega^{[1]}\right)_x=0,\\& \omega^{[1]}_x-uv^{[1]}_x=0,\end{aligned}\end{equation} 
which can be considered as a modified Camassa--Holm equation.

\item From \eqref{3.23} and \eqref{3.34}, it is easy to see that $\partial_2 X=\partial_2 x=-1$. Therefore, the reductions of \eqref{3.27} and \eqref{3.35} are
\begin{equation}
\partial_{0001}M+4\partial_1 M\,\partial_{00}M+8\partial_0 M\,\partial_{01}M=0,\end{equation} 
which is the AKNS equation, and
\begin{equation}
\partial_0\left(\frac{\partial_{001}x-1}{\partial_0 x}\right)=\partial_1\left(\frac{\partial_0 x^2}{2}\right),\end{equation} 
which is the modified AKNS equation.

\end{itemize}

\section{Conclusions}

Concerning the role of reciprocal transformations in the classification and identification of PDEs, we have shown that CH(2+1) and mCH(2+1) hierarchies can be connected with the CBS and mCBS equations via a reciprocal transformations. A big advantange of a reciprocal transformation is that it turns a whole hierarchy into a set of equations that can be studied through Painlev\'e analysis and other properties can afterwards be derived from this.

In this context, a reciprocal transformation has served as a way to turn a set of differential equations with multiple scalar fields a few independent variables into a unique differential equation with one scalar field depending on multiple independent variables. Furthermore, it serves to turn the initial equations into one in which the Painlev\'e integrability is satisfied and therefore proving the integrability of the hierarchy prior to the reciprocal transformation.

We have shown examples of higher-order by presenting a fourth-order nonlinear PDE in $2+1$ dimensions and investigated different reciprocal transformations
for it. Reciprocal transformations have once more shown that the transformed equations (their reductions actually) in $1+1$ dimensions are the Vakhnenko--Parkes and Degasperis-–Procesi equations.

Reciprocal transformations have been further proved to be useful for the derivation of Lax pairs. As it has been shown, the transformations of CH and mCH into CBS and mCBS, being these later equations integrable in the algebraic Painlev\'e sense, and being their Lax pair knowledgeable, undoing the reciprocal transformation in the Lax pairs for CBS and mCBS, we were able to retrieve Lax pairs that have not been proposed for CH and mCH in $1+1$ and $2+1$ dimensions. This verifies the importance of reciprocal transformations as a way to derive Lax pairs.

As a last instance, we have depicted Miura-reciprocal transformations, based on the composition of a Miura transformation between the CBS and mCBS and the reciprocal transformations linking CH and mCH to CBS and mCBS, correspondingly, in $1+1$ and $2+1$. Miura-reciprocal transformations verify the importance of composition of reciprocal transformations to classify hierarchies, indeed, we have successfully proven that CH and mCH in $1+1$ and $2+1$ are two different versions of a same common problem that can be reached by a transformation map that has been proposed in the last section.

The observation of all these properties show the efficiency and importance of reciprocal transformations that we introduced at the start of the chapter, and that we here close having given proof of our arguments with remarkable examples in the physics literature of hydrodynamic systems, shallow water waves, etc.


\begin{thebibliography}{99}

\bibitem{AbloClark}
Ablowitz M J and Clarkson P A, \textit{Solitons, nonlinear evolution equations and inverse
scattering}, London Mathematical Society, Lecture Notes Series 149, Cambridge
University Press, Cambridge, 1991.

\bibitem{AbloKruskalSegur}
Ablowitz M J, Kruskal M and Segur H, A note on Miura's transformation, \textit{J.
Math. Phys.} \textbf{20}, 999–-1003, 1979.

\bibitem{AbloSegur}
Ablowitz M J and Segur H, \textit{Solitons and the inverse scattering transform}, Society
for Industrial and Applied Mathematics (SIAM), Philadelphia, 1981.

\bibitem{Ab}
Abraham R and Marsden J E,
{\it Foundations of Mechanics}, 2nd edition,
Addison--Wesley, 1978.

\bibitem{bog}
Bogoyavlenskii O I, Breaking solitons in 2+1-dimensional integrable equations, \textit{Russian Math. Surveys} \textbf{45}, 1-86, 1990.

\bibitem{cal}
Calogero F,  A method to generate solvable nonlinear evolution equations, {\it Lettere al Nuovo Cimento}  \textbf{34}, 2443--447, 1975.

\bibitem{calogero} 
Calogero F,  Generalized Wronskian relations, one-dimensional Schr\"odinger equation and non-linear partial differential equations solvable by the inverse-scattering method, {\it Nuovo Cimento B}  \textbf{31}, 229-249, 1976.



\bibitem{CH}
Camassa R and Holm D D, An integrable shallow water equation with peaked solitons, \textit{Phys. Rev. Lett.}, \textbf{71}(11), 1661–-1664, 1993.

\bibitem{clarkson}
Clarkson P A, Fokas A S and Ablowitz M J, Hodograph transformations
of linearizable partial differential equations, {\it SIAM J. of Appl. Math.} {\bf 49},
1188–-1209, 1989.

\bibitem{const2}
Constantin A, On the scattering problem for the Camassa–Holm equation, \textit{R. Soc. Lond. Proc. Ser. A Math. Phys. Eng. Sci.}, \textbf{457}, 2001

\bibitem{const3}
Constantin A, Gerdjikov V S and Ivanov R I, Inverse scattering transform for the Camassa–Holm equation, \textit{Inverse Problems}, \textbf{22}(6), 2197–-2207, 2006

\bibitem{const1}
Constantin A and McKean H P, A shallow water equation on the circle, \textit{Commun. Pure Appl. Math.}, \textbf{52}(8), 949–-982, 1999

\bibitem{conte}
Conte R and Musette M, {\it The Painlev\'e Handbook}, Springer and Canopus Publishing
Limited, Bristol, 2008.

\bibitem{degasholm}
Degasperis A, Holm D D and Hone A N W, A new integral equation with
peakon solutions, {\it Theor. Math. Phys.} {\bf 133}, 1463–-1474, 2002.


\bibitem{Est1}
Est\'evez P G, Reciprocal transformations for a spectral problem in 2 + 1
dimensions, \textit{Theor. and Math. Phys.} \textbf{159}, 763–-769, 2009.

\bibitem{Est2}
Est\'evez P G, Generalized Qiao hierarchy in 2+1 dimensions: reciprocal
transformations, spectral problem and non-isospectrality, \textit{Phys. Lett. A},
\textbf{375}, 537--540, 2011.

\bibitem{EstGand}
Est\'evez P G, Gandarias M L and Prada J, \textit{Phys. Lett. A}, Symmetry reductions of a 2+1 Lax pair \textbf{343}, 40–-47, 2005.

\bibitem{EstLeble}
Est\'evez P G and Leble S L, A wave equation in 2+1: Painleve analysis and solutions, \textit{Inverse Problems}, \textbf{11}, 925–-937, 1995.

\bibitem{EstPrada}
Est\'evez P G and Prada J, A Generalization of the Sine-Gordon Equation to
2 + 1 Dimensions, \textit{J. of Nonlinear Math. Physics} \textbf{11}, 164-179, 2004.

\bibitem{estevez05-1} 
Est\'evez P G and Prada J, Hodograph Transformations for a Camassa--Holm hierarchy in 2+1 dimensions, \textit{J. Phys. A: Math. Gen.} \textbf{38}, 1287--1297, 2005.

\bibitem{EstPra2}
Est\'evez P G and Prada J Singular Manifold Method for an Equation in $2+1$ Dimensions, {\it J. Nonlin. Math. Phys} {\bf 12}, 266--279, 2005.

\bibitem{EstSar2}
Est\'evez P G and Sard\'on C, Miura reciprocal transformations for two integrable hierarchies in 1 + 1 dimensions, \textit{Proceedings GADEIS (2012)}, Protaras, Cyprus, 2012.

\bibitem{EstSar1}
Est\'evez P G and Sard\'on C, Miura reciprocal Transformations for hierarchies in 2+1 dimensions, \textit{J. Nonlin. Math. Phys.} \textbf{20}, 552–-564, 2013.


\bibitem{Ferapontov1}
Ferapontov E V, Reciprocal transformations and their invariants, \textit{Differential
equations} \textbf{25}, 898–-905, 1989.

\bibitem{FerapontovRogersSchief}
Ferapontov E V, Rogers C and Schief W K, Reciprocal transformations of two
component hyperbolic system and their invariants, \textit{J. Math. Anal. Appl.} \textbf{228},
365–-376, 1998.

\bibitem{h00} 
Hone A N W,  Reciprocal link for $2+1$-dimensional extensions of shallow water equations,
{\it App. Math. Letters} {\bf 13}, 37--42, 2000.

\bibitem{ivanov} 
Ivanov R, Equations of the Camassa--Holm hierarchy, \textit{Theor. and Math. Phys.}, \textbf{160}, 953--960, 2009.
 
 \bibitem{pick}  
 Kudryashov N  and Pickering A, Rational solutions for Schwarzian integrable hierarchies, \textit{J. Phys. A: Math. Gen.} {\bf 31}, 9505--9518, 1998.
 
 \bibitem{kruskal}
 Kruskal M, Nonlinear Wave Equations, In Moser J, {\it Dynamical Systems, Theory and Applications} {\bf 38}, 310--354, Springer, 1975.
 
 \bibitem{oevelrogers}
Oevel W and Rogers C, Gauge transformations and reciprocal links in
2 + 1 dimensions, \textit{Rev. Math. Phys.} \textbf{5}, 299–-330, 1993.
 
 \bibitem{HolmQiao}
Qiao Z, The Camassa-–Holm hierarchy, related N-dimensional integrable systems, and algebro--geometric solutions on a symplectic submanifold, {\it Commun. Math. Phys.} {\bf 239}, 309--341, 2003.
 
 
\bibitem{qiao}
Qiao Z, A new integrable equation with cuspons and W/M--shape--peaks solitons, \textit{J. Math. Phys.} \textbf{47}, 112701, 2006. 
 
 \bibitem{qiao2007} 
 Qiao Z, New integrable hierarchy, its parametric solutions, cuspons, one--peak solitons, and M/W--shape solitons, \textit{J. Math. Phys.} \textbf{48}, 082701, 2007.

\bibitem{QiaoLiu}
Qiao Z and Liu L, A new integrable equation with no smooth solitons, \textit{Chaos, Solitons and Fractals} \textbf{41}, 587, 2009.

 \bibitem{Rogers11}
Rogers C, Application of a reciprocal transformation to a two-phase Stefan
Problem, \textit{J. Phys. A} \textbf{18}, L105--L109, 1985.
 
 \bibitem{rogers4}  
 Rogers C, Reciprocal transformations in (2+1) dimensions, {\it J. Phys. A: Math. Gen} \textbf{19}, L491-L496, 1986.

\bibitem{rogers5}  
Rogers C, The Harry Dym equation in 2+1 dimensions: A reciprocal link with the Kadomtsev--Petviashvili equation, {\it Phys. Lett. A} \textbf{120}, 15--18, 1987.

\bibitem{rogerscarrillo}
Rogers C and Carillo S, On reciprocal properties of the Caudrey, Dodd–-
Gibbon and Kaup–-Kuppersmidt hierarchies, \textit{Phys. Scripta} \textbf{36}, 865-–869,
1987.

\bibitem{RogersKingstonShadwick}
Rogers C and Kingston J G and Shadwick W F,  On reciprocal type invariant transformations
in magneto-gas dynamics, \textit{J. Math. Phys.} \textbf{21}, 395–-397, 1980.

\bibitem{rogersnucci}
 Rogers C and Nucci M C, On reciprocal Bäcklund transformations and
the Korteweg de Vries hierarchy, \textit{Phys. Scripta} 33, 289–292, 1986.

\bibitem{rogersshadwick}
Rogers C and  Shadwick W F, \textit{B\"acklund transformations and their applications}, Mathematics in scince and engineerin, Vol 161,  Academic Press, 1982.

\bibitem{Sakovich}
Sakovich S Y, On Miura transformations of evolution equations, \textit{J. Phys. A:
Math. Gen.} \textbf{26}, L369–-L373, 1993.

\bibitem{vakh}
Vakhnenko V O, Solitons in a nonlinear model medium, \textit{J. Phys. A}, \textbf{25}, 4181-–4187, 1992.

\bibitem{weiss}
Weiss J, The Painlev\'e property for partial differential equations. II: B\"{a}cklund transformation, Lax pairs, and the Schwarzian derivative, \textit{J. Math. Phys.}, \textbf{24}, 1405-–1413, 1983.

\bibitem{ZS}
Zakharov V E and Shabat A B, Exact theory of two dimensional self-focusing
and one-dimensional self-modulation of waves in nonlinear media, \textit{Sov. Phys.
JETP}, \textbf{34}, 62–-69, 1972.


\end{thebibliography}
\end{document}